\documentclass[aps, prb, notitlepage, citeautoscript,preprintnumbers, superscriptaddress, eqsecnum, reprint]{revtex4-2}

\usepackage{xr}
\usepackage{amsmath}
\usepackage{amsfonts, amssymb,amsxtra}
\usepackage[]{graphicx} 
\usepackage{grffile}
\usepackage{amsfonts}
\usepackage{framed}
\usepackage{bbm}
\usepackage{braket}
\usepackage{xcolor}
\usepackage{mathtools}
\usepackage{diagbox}
\usepackage{LatexCommands}
\usepackage{tabularx}
\usepackage{tikz}
\usetikzlibrary{quantikz}
\newcommand{\tr}{\text{Tr}}

\usepackage[colorlinks=true]{hyperref}
\hypersetup{
    unicode=false,          % non-Latin characters
    pdftoolbar=true,        % show Acrobat
    pdfmenubar=true,        % show Acrobat
    pdffitwindow=false,     % window fit to page when opened
    pdfstartview={FitH},    % fits the width of the page to the window
    pdftitle={},    % title
    pdfauthor={},     % author
    pdfsubject={},   % subject of the document
    pdfcreator={},   % creator of the document
    pdfproducer={}, % producer of the document
    pdfkeywords={} {} {}, % list of keywords
    pdfnewwindow=true,      % links in new window
    colorlinks=true,       % false: boxed links; true: colored links
    linkcolor=magenta, %red,          % color of internal links (change box color with linkbordercolor)
    citecolor=blue,        % color of links to bibliography
    filecolor=magenta,      % color of file links
    urlcolor=blue           % color of external links
}

\begin{document}
\preprint{FERMILAB-PUB-23-268-SQMS-T}
%\title{Mixed field Ising model dynamics using pulse level controlled Trotter quantum circuits}
\title{Reconstructing Thermal Quantum Quench Dynamics from Pure States}

\author{Jason Saroni}
\email{jsaroni@iastate.edu}
\affiliation{Superconducting Quantum Materials and Systems Center (SQMS), Fermi National Accelerator Laboratory, Batavia, Illinois 60510, USA}
\affiliation{Ames National Laboratory, Ames, Iowa 50011, USA}
\affiliation{Department of Physics and Astronomy, Iowa State University, Ames, Iowa 50011, USA}

\author{Henry Lamm}
\email{hlamm@fnal.gov}
\affiliation{Superconducting Quantum Materials and Systems Center (SQMS), Fermi National Accelerator Laboratory, Batavia, Illinois 60510, USA}
\affiliation{Fermi National Accelerator Laboratory, Batavia, Illinois, 60510, USA}

\author{Peter P.~Orth}
\email{peter.orth@uni-saarland.de}
\affiliation{Superconducting Quantum Materials and Systems Center (SQMS), Fermi National Accelerator Laboratory, Batavia, Illinois 60510, USA}
\affiliation{Ames National Laboratory, Ames, Iowa 50011, USA}
\affiliation{Department of Physics and Astronomy, Iowa State University, Ames, Iowa 50011, USA}
\affiliation{Department of Physics, Saarland University, 66123 Saarbr\"ucken, Germany}

\author{Thomas Iadecola}
\email{iadecola@iastate.edu}
\affiliation{Superconducting Quantum Materials and Systems Center (SQMS), Fermi National Accelerator Laboratory, Batavia, Illinois 60510, USA}
\affiliation{Ames National Laboratory, Ames, Iowa 50011, USA}
\affiliation{Department of Physics and Astronomy, Iowa State University, Ames, Iowa 50011, USA}

\begin{abstract}
Simulating the nonequilibrium dynamics of thermal states is a fundamental problem across scales from high energy to condensed matter physics. Quantum computers may provide a way to solve this problem efficiently. Preparing a thermal state on a quantum computer is challenging, but there exist methods to circumvent this by computing a weighted sum of time-dependent matrix elements in a convenient basis. While the number of basis states can be large, in this work we show that it can be reduced by simulating only the largest density matrix elements by weight, capturing the density matrix to a specified precision. Leveraging Hamiltonian symmetries enables further reductions. This approach paves the way to more accurate thermal-state dynamics simulations on near-term quantum hardware.
\end{abstract}
\date{\today}

\maketitle

\section{Introduction}
\label{sec:Introduction}

Simulating quantum quench dynamics is a classically hard problem due to the generation of entanglement in the post-quench nonequilibrium state. It is thus a natural application for quantum computers~\cite{Feynman1982,doi:10.1126/science.273.5278.1073, miessenQuantumAlgorithmsQuantum2023}, which do not suffer from the exponential increase of resources that classical simulations experience~\cite{PhysRevE.75.015202}. A quench is a process in which a model parameter of a quantum system changes abruptly in time, taking a stationary state into a complex superposition of excited states~\cite{barouchStatisticalMechanicsMathrmXY1970, igloiLongRangeCorrelationsNonequilibrium2000, senguptaQuenchDynamicsQuantum2004, barankovCollectiveRabiOscillations2004, calabreseQuantumQuenchesExtended2007, kollathQuenchDynamicsNonequilibrium2007b, moeckelInteractionQuenchHubbard2008b, PhysRevB.81.012303, PhysRevLett.106.227203,gagelUniversalPostquenchCoarsening2015, PhysRevLett.113.220401,   cuiPostquenchGapDynamics2019a, annurev-conmatphys-031016-025451, acrefore/9780190871994.013.55}.
%Li2009ExactRF,
Quench dynamics are of interest across physics, from cold atomic gases trapped in optical lattices~\cite{Greiner2002, Kinoshita2006,  RevModPhys.80.885, Bloch2012, gringRelaxationPrethermalizationIsolated2012b, langenExperimentalObservationGeneralized2015, nicklasObservationScalingDynamics2015b, langenUltracoldAtomsOut2015b, PhysRevLett.121.250403} to ultrafast pump-probe experiments of solid-state systems~\cite{ kampfrathResonantNonresonantControl2013b, Li,matsunagaLightinducedCollectivePseudospin2014a, yangTerahertzlightQuantumTuning2018a} and hadronization in heavy-ion collisions~\cite{ARSENE20051}. Often quenches from pure states are studied, but starting from initially mixed states like thermal states is also physically interesting and experimentally relevant~\cite{PhysRevE.75.015202, Sotiriadis2009QuantumQF, PhysRevB.93.104302}. 

Quenches from pure states can be simulated naturally on quantum computers using, e.g. Suzuki-Trotter decomposition \cite{Trotter1959,Suzuki1976,doi:10.1126/science.273.5278.1073} or variational approaches \cite{s42254-021-00348-9, Yuan2019, PRXQuantum.2.030307} (see Ref.~\onlinecite{miessenQuantumAlgorithmsQuantum2023} for a recent overview on quantum computing simulation methods for quantum dynamics). However, the evolution of mixed, thermal, or thermofield double~\cite{Maldacena2003} states encounters the challenge of preparing such states on a quantum computer~\cite{Motta2020, wu2019a}.
One means of circumventing this costly state preparation is the ``evolving density matrices on qubits" (E$\rho$OQ) algorithm~\cite{Lamm:2018siq,Harmalkar:2020mpd,Gustafson:2020yfe}. E$\rho$OQ is a hybrid quantum-classical algorithm in which the initial density matrix $\rho$ in some computationally simple basis of $N$ states is obtained classically using a stochastic method such as density matrix quantum Monte Carlo (DMQMC)~\cite{PhysRevB.89.245124} or Euclidean lattice field theory. The quantum dynamics simulations are then performed with corresponding basis states which can be more easily prepared, and the convolution with the classically obtained $\rho$ reconstructs the full mixed-state dynamics.

While E$\rho$OQ is an appealing hybrid algorithm, it potentially requires a unique quantum simulation for all $N^2$ elements of $\rho$ to reconstruct the dynamics. However, often it may be unnecessary to evolve all initial states; e.g., in gapped systems at a low temperature, only a limited set of configurations may be required.  Further, one can truncate $\rho$ to reduce the number of quantum simulations to $N_{\rm sim} < N^2$ at the price of systematically-improvable errors.

Here we study the effect of truncation on the simulated dynamics of thermal states. We investigate the structure of thermal $\rho$ and simulate the dynamics of such states with varying $N_{\rm sim}$. Optimizations of the truncation approximation are investigated, such as choosing appropriate bases for $\rho$ depending on the model parameters and the observables measured. Another optimization is to exploit symmetry relations between basis states to reduce the number of unique simulations required. In doing so, we demonstrate that $N_{\rm sim}$ can be reduced by up to two orders of magnitude from the naive estimate. We apply our techniques to quenches from thermal states in the mixed-field Ising model, %\cite{PhysRevB.68.214406}
using both exact diagonalization (ED) and DMQMC to prepare $\rho$. While our study is inspired by quantum computing as a promising use case, we carry it out on classical computers as we are addressing questions of principle rather than implementation. The latter direction is a logical next step for future work, and we briefly comment further on it in our conclusion.

The remainder of the paper is organized as follows. First, we briefly review the E$\rho$OQ and DMQMC algorithms in Sec.~\ref{sec:epoq}. In Sec.~\ref{sec:model_and_quench_protocol}, the mixed-field Ising model and the quench protocol are defined. Sec.~\ref{structure_and_truncation_of_the_initial_density_matrix} is devoted to the impact of truncating $\rho$. This is followed by Sec.~\ref{sec:leveraging_symmetries} where symmetries are used to further reduce $N_{\rm sim}$. Classical dynamics simulations are presented in Sec.~\ref{Results} to demonstrate how accuracy depends on the truncation. We consider one system size where $\rho$ is easily accessible by ED, and another where full ED becomes expensive and DMQMC becomes a desirable algorithmic choice. Sec.~\ref{Discussion_and_outlook} concludes and describes future steps.

\section{E\texorpdfstring{$\rho$}{p}OQ algorithm}
\label{sec:epoq}
Ultimately, our goal is to compute the expectation value of an observable $O$ at a time $t$. In the Heisenberg picture, the time evolution of $O$ with a Hamiltonian $H_1$ is given by $O(t)=e^{iH_1t}Oe^{-iH_1t}$.  Our initial state is defined via a Hamiltonian $H_0$ by $\rho=e^{-\beta H_0}/\tr(e^{-\beta H_0})$ and has an inverse temperature $\beta=1/T$.   Then, the expectation value is
\begin{equation}
\label{eq:O(t)}
\braket{O(t)}_{\rho}
=
\text{Tr}[\rho O(t)]
=
\sum_{m,n}
\braket{m|\rho|n}
\braket{n|O(t)|m} 
\end{equation}
where we have resolved the trace in a complete basis.

The E$\rho$OQ algorithm of Ref.~\cite{Lamm:2018siq} reconstructs the dynamics in Eq.~\eqref{eq:O(t)} by first classically obtaining a stochastic approximation $\tilde{\rho}$ and the matrix elements $\tilde{\rho}_{mn}=\braket{m|\tilde{\rho}|n}$ using DMQMC~\cite{PhysRevB.89.245124}. With a stochastic algorithm like DMQMC, it is possible to reach larger system sizes than ED. The un-normalized thermal $\tilde{\rho}(\beta) = e^{-\beta H_0} $ is approximated through DMQMC by a stochastic solution to the symmetric Bloch equation 
\begin{align}
\label{eq:sbe}
\frac{\mathrm{d} \tilde{\rho}}{\mathrm{d} \beta} = -\frac{1}{2}( H_0\tilde{\rho}+ \tilde{\rho} H_0 )
\end{align}
with the initial condition $\tilde{\rho}(\beta=0) = \mathbbm{1}$~\cite{Lamm:2018siq}.  This stochastic solution is obtained in DMQMC by discretizing $\beta=N_{\beta}\delta\beta$.  Then one initializes a number $N_{\rm psip}$ of imaginary particles called ``psips" in the diagonal states $|n\rangle\langle n|$. At each $\delta\beta$ step, these psips are allowed to move in the space of basis states $|m\rangle\langle n|$ with probabilistic rules derived from Eq.~\eqref{eq:sbe}. Note that each psip also carries a sign, so that its contribution to a given density matrix element can be positive or negative.  From this, $\tilde{\rho}(\beta)$ is obtained as a sum over psips:
\begin{align}
\tilde{\rho}(\beta) = \frac{1}{2\chi_{\rm diag}}\sum_{mn} \left(  \chi_{mn} \ket{m}\bra{n}+ \chi_{mn}^* \ket{n}\bra{m}\right),
\label{eq:rho_dmqmc}
\end{align}
where $\chi_{mn}$ is determined by the number and sign of the psips associated with $\ket{m}\bra{n}$ and $\chi_{\rm diag} = \sum_i \chi_{ii}$ ensures normalization. (We note that $\chi_{mn}$ are taken in this work to be real.) The finiteness of $N_{\rm psip}$ naturally truncates $\rho$ since any matrix element with $|\rho_{mn}|<1/N_{\rm psip}$ will be zero. The statistical error of $\tilde{\rho}$ is derived from Poisson statistics in Appendix~\ref{sec:statistical_error}. This error can be systematically reduced by including more psips in DMQMC.

With the classical simulation (using either DMQMC or ED) designating which $\rho_{mn}$ are nonzero, the quantum computer is used to obtain the matrix elements $O_{nm}=\braket{n|O(t)|m}$. Combining these with $\tilde{\rho}_{mn}$ produces an approximation to Eq.~\eqref{eq:O(t)}. To obtain the matrix elements $O_{nm}$, one initializes the quantum computer in the superposition states
\begin{align}
\label{eq:purepsiphi}
\begin{split}
\left | \psi^{\pm}_{nm} \right \rangle &= \frac{1}{\sqrt{2}}\left( \left | n \right \rangle \pm \left | m \right \rangle \right )\\
\left | \phi^{\pm}_{nm} \right \rangle &= \frac{1}{\sqrt{2}}\left( \left | n \right \rangle \pm i\left | m \right \rangle \right )
\end{split}
\end{align}
and evaluates the time-dependence of $O$ via
%\begin{widetext}
\begin{align}
\begin{split}
\label{eq:Onm}
\text{Re}\,O_{nm} &= \frac{1}{2}\left( \langle \psi^{+}_{nm} |  O(t) | \psi^{+}_{nm} \rangle - \langle \psi^{-}_{nm} |  O(t) | \psi^{-}_{nm} \rangle  \right )\\
\text{Im}\,O_{nm}&= \frac{1}{2}\left( \langle \phi^{-}_{nm} |  O(t) | \phi^{-}_{nm} \rangle - \langle \phi^{+}_{nm} |  O(t) | \phi^{+}_{nm} \rangle  \right ).
\end{split}
\end{align}
%\end{widetext}
(We discuss the contribution of state preparation to the quantum simulation overhead in Appendix~\ref{sec:initial_state_preparation}.) Note that in many cases, $O_{nm}$ is purely real.
In any case, naively a number of quantum dynamics simulations $N_{\rm sim}\approx 
%\propto 
N^2$ is required to perfectly reconstruct the dynamics of a $\tilde{\rho}$ of size $N\times N$.

The goal of this work is to investigate whether $N_{\rm sim}$ can be further reduced by judicious truncation of $\tilde{\rho}$ and by leveraging symmetries. To address this question, we use classical simulations to obtain either $\rho$ itself by ED or $\tilde{\rho}$ by DMQMC. While the former approach only works for relatively small systems, it allows for faithful benchmarking against the exact dynamics.

\section{Model and quench protocol}
\label{sec:model_and_quench_protocol}

\begin{figure}[htb!]
\includegraphics[width=\linewidth]{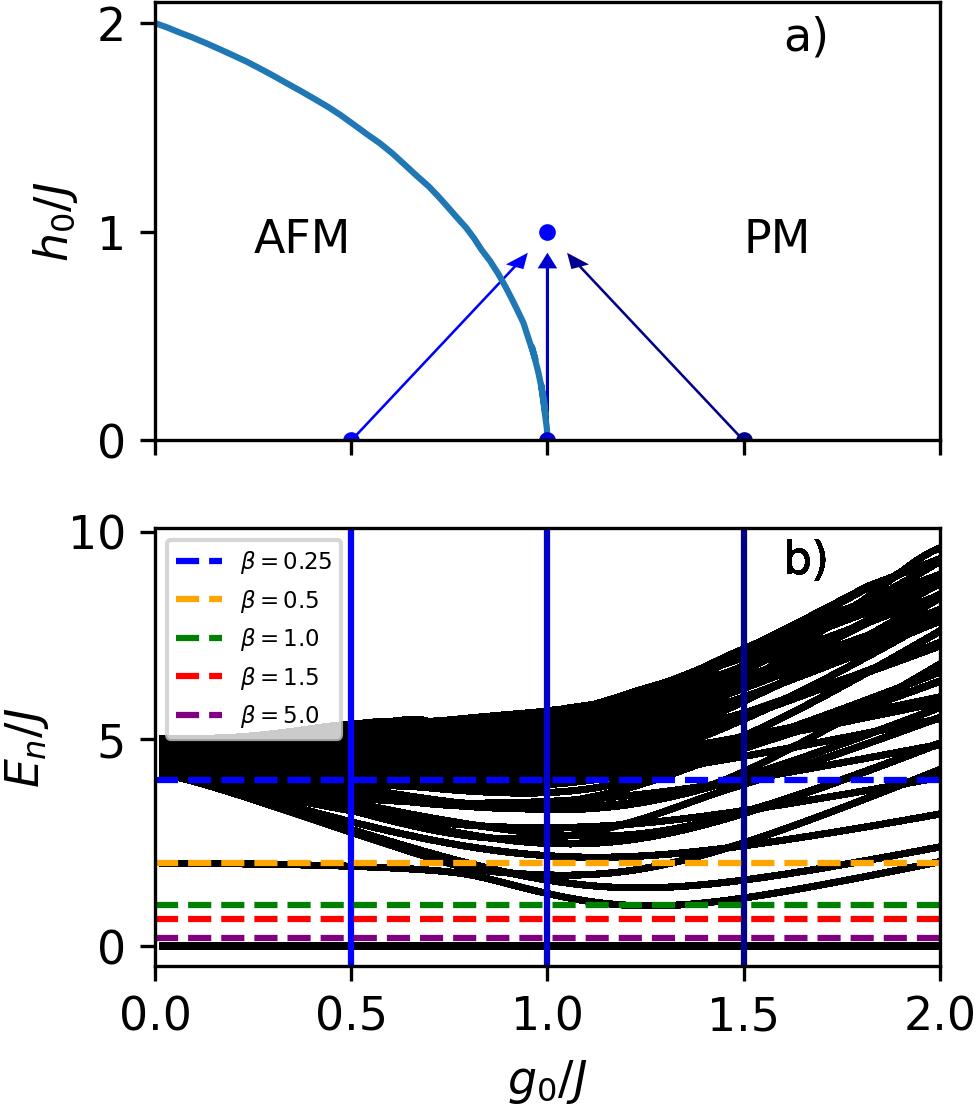} 
\caption{a) $T=0$ phase diagram of the MFIM with $h_s=0$ from Ref.~\cite{PhysRevB.68.214406}. The blue arrows of different shades show the quenches initial and final location at $h=g=1$. b) Lowest $100$ energy levels as a function of $g_0$ for fixed $h_0 = 0$, $h_s = 1/L$ and $J=1$ and $L=14$. $E_n$ denotes the $n$th energy eigenvalue of $H_0$ from lowest to highest. The dashed horizontal lines indicate $T=1/\beta$. The vertical blue lines correspond to the quenches in a).} 
\label{fig:spectrum_phase_diagram_H0}
\end{figure}

% The purpose of this section is to describe the Hamiltonian model and quenches used along different regions of its phase diagram. 
In this section we discuss the one-dimensional antiferromagnetic mixed-field Ising model (MFIM) and the specific quench protocol used in our simulations. The MFIM Hamiltonian $H$ is
\begin{align}
H &= \sum_{i=1}^{L} \Bigl[ JZ_i Z_{i+1} + g X_i + h Z_i +h_s (-1)^i Z_i \Bigr] \,,
\label{eq:H0}
\end{align}
and includes a nearest-neighbor antiferromagnetic (AFM) Ising coupling $J > 0$ and transverse and longitudinal fields $g$ and $h$.
We also include a small staggered magnetic field $h_s = L^{-1}$ to weakly lift the degeneracy of the ground state in the AFM phase. We assume periodic boundary conditions such that $Z_{L+1} \equiv Z_1$, and hereafter set $J=1$ as a unit of energy and inverse time.
We restrict to even $L$ for convenience.

At $T=0$ and $h_s=0$, the model exhibits AFM and paramagnetic (PM) phases separated by a continuous phase transition except at $g=0$, where the model reduces to the classical Ising model and exhibits a first-order transition at $h=2$.
The order parameter for the AFM phase is the staggered magnetization density
\begin{align}
M^z_\pi = \frac{1}{L}\sum^L_{i=1} (-1)^i Z_i.
\end{align}
In the disordered PM phase, as long as $g\neq 0$ the ground state acquires finite magnetization density
\begin{align}
M^x=\frac{1}{L}\sum^L_{i=1} X_i.
\end{align}
The $T=0$ phase diagram is found in Fig.~\ref{fig:spectrum_phase_diagram_H0}a) as determined by density matrix renormalization group (DMRG) simulations in Ref.~\cite{PhysRevB.68.214406}.

Although no ordered phase exists when $T\neq 0$, for sufficiently small $T$, $\rho$ has high weight on the N\'eel state $\ket{0101\dots}$ for parameters $g,h$ belonging to the zero-temperature AFM phase. Since the initial $T$ sets an energy scale below which states contribute significantly to $\rho$, increasing $T$ results in more energy eigenstates contributing to $\rho$, as shown in Fig.~\ref{fig:spectrum_phase_diagram_H0}b). Similarly, more states contribute to $\rho$ near the $L\to\infty$ critical point $g=J$, where the energy gap approaches its minimum. Thus, the structure of $\rho$ depends on parameters of the model and initial state, as we discuss in Sec.~\ref{structure_and_truncation_of_the_initial_density_matrix}.  

In our simulations, $H_0$ and $H_1$ are given by the MFIM Hamiltonian~\eqref{eq:H0} with different choices of couplings. 
$H_0$ is defined by the couplings $g_0,h_0,h_s$, while the quench Hamiltonian $H_1$ is defined by couplings $g$ and $h$ with $h_s=0$.
Physically, this corresponds to an abrupt parameter quench in which the fields are changed: $g_0\rightarrow g$, $h_0\rightarrow h$, and $h_s\rightarrow 0$.
Eq.~\eqref{eq:H0} is nonintegrable for $h\neq 0$~\cite{PhysRevE.90.052105, PhysRevE.75.015202}.
We consider three quenches to the point $h/J=g/J=1$, starting from the initial points $h_0/J=0$ and $g_0/J=0.5,1.0,1.5$. These quenches are represented by arrows in Fig.~\ref{fig:spectrum_phase_diagram_H0}a).

\section{Structure and Truncation of \texorpdfstring{$\rho$}{p}}
\label{structure_and_truncation_of_the_initial_density_matrix}

\begin{figure}[t!]
\includegraphics[width=\linewidth]{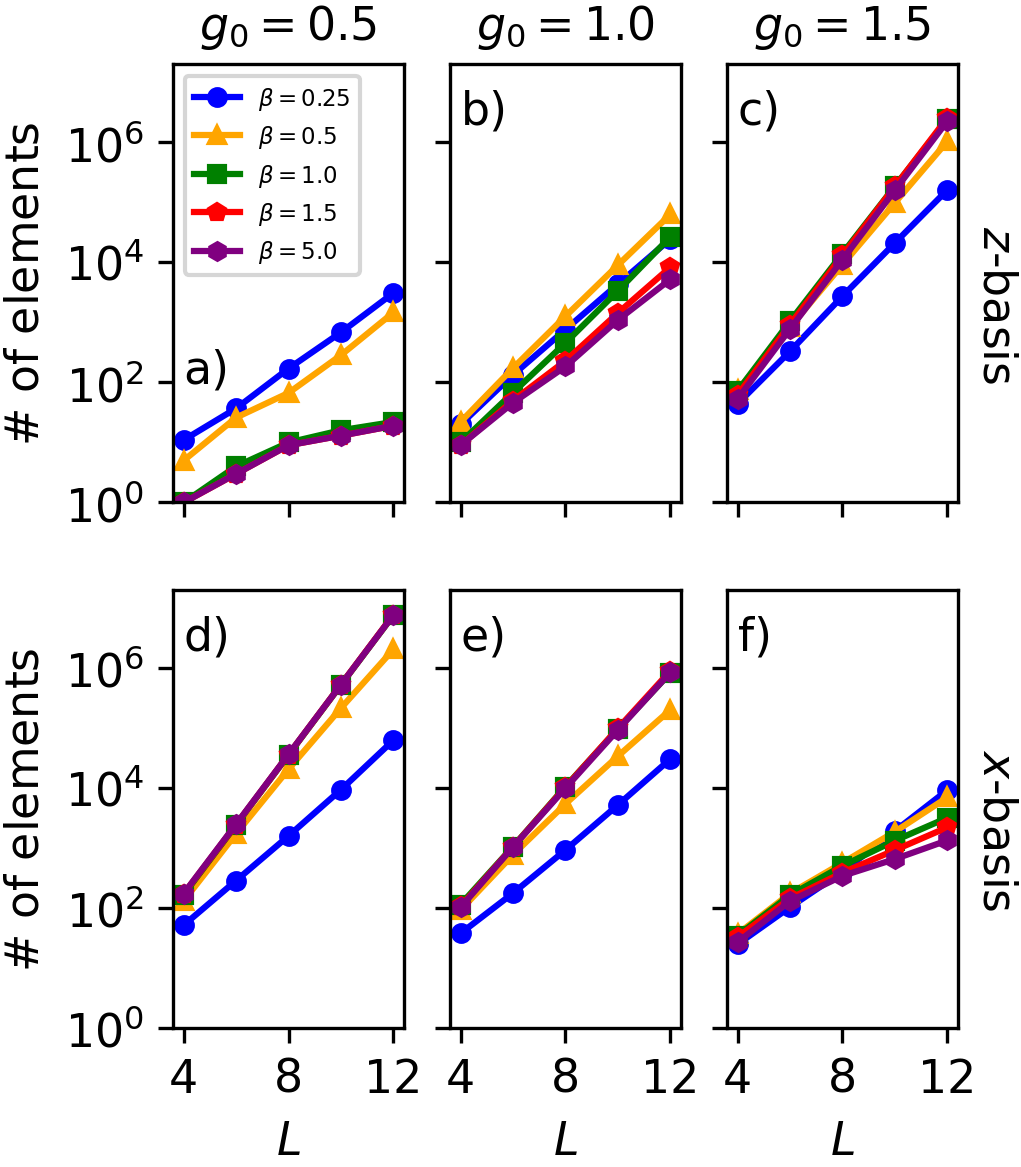}
\caption{
$L$ dependence of the number of elements of the truncated density matrix $\rho^w$ needed to reach $w\approx 0.93$ [see Eq.~\eqref{eq:weight}] for initial state parameters $g_0=0.5,1.0,$ and $1.5$ with $h_0=0$. A range of temperatures is plotted with $\rho^w$ represented in Panels a)--c) in the $z$-basis and in Panels d)--f) in the $x$-basis.
}
\label{fig:weight_plots}
\end{figure}

To reduce the simulation cost of evaluating Eq.~\eqref{eq:O(t)}, we now investigate how to systematically omit some $\rho_{mn}$ while retaining a certain accuracy. As mentioned in Sec.~\ref{sec:Introduction}, we expect that certain parameter regimes of $H_0$ should be less sensitive to discarding specific $\rho_{mn}$. For example, in the AFM phase at large $\beta$, $\rho$ has high weight on only a few $\rho_{mn}$: specifically, if $\rho$ is represented in the $z$-basis, the N\'eel states $\ket{10\dots}$ and $\ket{01\dots}$ should dominate over all other configurations. Similarly, if the initial state is deep in the PM phase, a single spin-polarized configuration should dominate. In contrast, for initial states close to the phase boundary---where the energy gap scales as $1/L$---or at small $\beta$, we expect many large $\rho_{mn}$. 

To make this intuition quantitative, we consider a truncated density matrix $\rho^w$ with $N_w$ nonzero elements as a function of $\beta$ and couplings. $\rho^w$ is defined via a set of indices $\mathcal{W} = \{(n,m) : |\rho_{nm}| > \epsilon \}$, with $\rho^w_{nm} \equiv 0$ for all $(n,m) \notin \mathcal{W}$ (note that $|\mathcal W|=N_w$).  The cutoff $\epsilon > 0$ is chosen such that the weight $w$, defined via the ratio of Frobenius norms
\begin{equation}
\label{eq:weight}
 w^2 = \frac{\left \| \rho^w  \right \|^2_F}{\left \| \rho  \right \|^2_F} = \frac{\text{Tr}\left (\rho^w\rho^{w\dagger} \right )}{\text{Tr}\left (\rho \rho^{\dagger} \right )}, 
 % = \frac{ \sum_{i,j=1}^{N^{\prime}}\left| \rho^{w}_{ij} \right|^2 }{ \sum_{i,j=1}^{N}\left| \rho_{ij} \right|^2  },
 \end{equation}
is above a fixed threshold.
%where $N$ for the MFIM is $2^L$, and $N^{\prime}$ are upper bounds on the indices of $\rho^w$ due to truncation. 
Note that we normalize the truncated density matrix $\rho^w \rightarrow \rho^w/\text{Tr}(\rho^w)$.

In Fig.~\ref{fig:weight_plots}, we calculate the $N_w$ required to achieve $w=0.93$ as a function of $\beta$ and system size $L$ for each quench $g_0=0.5,1.0,1.5$ with $h_0=0$. Since this quantity depends on the basis in which $\rho^w$ is represented, we show results for both the $z$- and $x$-basis. Focusing first on the $z$-basis results, there is a clear trend of increasing $N_w$ with increasing $g_0$. This is due to the fact that the initial state becomes polarized in the $x$-basis for large $g_0$, and therefore can only be represented using all $z$-basis states. For example, in the extreme limit $g_0\to \infty$ and $\beta\to \infty$, $\rho\propto (\ket{-}\bra{-})^{\otimes L}$ becomes fully dense, involving all $2^{2L}$ matrix elements in the $z$-basis. Remnants of this behavior are clearly visible for the larger $\beta$ values in Fig.~\ref{fig:weight_plots}c) ($g_0=1.5$), where the required $N_w$ approaches $2^{2L}$. As $\beta$ decreases, $\rho$ becomes more diagonal, so that at $\beta=0$ only $2^L$ matrix elements are nonzero. This effect is also visible in the $\beta=0.25$ curve in Fig.~\ref{fig:weight_plots}c). Meanwhile, in the AFM phase at $g_0=0.5$ [Fig.~\ref{fig:weight_plots}a)], we see that far fewer $z$-basis matrix elements are required, especially at large $\beta$. However, in all phases we observe the general tendency that $N_w$ increases exponentially with $L$.

The $x$-basis results mirror the $z$-basis results. In the AFM phase [Fig.~\ref{fig:weight_plots}d), $g_0=0.5$], where the density matrix is sparser in the $z$-basis, the $x$-basis representation of $\rho^w$ is much denser. In contrast, the density matrix sparsens in the PM phase [Fig.~\ref{fig:weight_plots}f), $g_0=1.5$], especially at high $\beta$. Finally, we observe that near the $T=0$ quantum critical point [Fig.~\ref{fig:weight_plots}b) and e), $g_0=1.0$] the difference in $N_w$ between different basis representations is greatly reduced, especially at low $\beta$. This is consistent with the intuition that more density matrix elements should be required in any basis near the phase transition due to the pileup of eigenstates at low energies.

\section{Leveraging Symmetries}
\label{sec:leveraging_symmetries}

In general, $\rho_{mn}$ and $O_{nm}$ are constrained by symmetries of the Hamiltonians $H_0$ and $H_1$. Here we discuss how to leverage these to further reduce $N_{\rm sim}$.

The symmetries of $H_0$ [Eq.~\eqref{eq:H0}] impose a degeneracy structure on $\rho$.
$H_0$ has a two-site translation symmetry owing to the presence of $h_s$; we denote the generator of this symmetry by $T_2$. Additionally, when $h_0,h_s=0$ the model is invariant under the global spin-flip $S=\prod_i X_i$. This operation commutes with the $ZZ$ and $X$ terms in $H_0$ but anticommutes with the staggered field. However, the latter also anticommutes with bond-centered reflection $R$, which maps site $i$ to site $L-i+1$ when $L$ is even, and one-site translation $T_1$, both of which commute with the remainder of $H_0$. Thus, the combined operations $SR$, $ST_1$, and $RT_1$ are symmetries for $h_0=0$. The impact of these symmetries on the structure of $\rho$ can be seen as follows.  Given a unitary operator $Q$ that commutes with $H_0$, any $\rho_{mn}\propto (e^{-\beta H_0})_{mn}$ satisfies
\begin{align}
\label{eq:rho_0-trafo}
\braket{m|\rho|n} = \braket{m|Q^\dagger\rho Q|n}=\braket{m'|\rho|n'},
\end{align}
where $\ket{m'}=Q\ket{m}$ and $\ket{n'}=Q\ket{n}$. Thus, given $\rho_{mn}$ in any basis, one immediately knows $\rho_{m'n'}$. Since $(ST_1)^L= (RT_1)^2=\mathbbm 1$ and $(ST_1)^2=T_2$, we can obtain from any $\rho_{mn}$ at most $4L-1$ additional elements. This is particularly useful when using DMQMC, which does not \emph{a priori} preserve the symmetries of $\rho$. By demanding that symmetry-related $\rho_{mn}$ are identical, we can produce a symmetrized DMQMC estimate $\tilde{\rho}$.

The symmetries of $H_1$, namely $T_1$ and $R$, can also be used to reduce $N_{\rm sim}$. The effect of these symmetry transformations on $O_{nm}$ depends on the choice of observable $O$. For example, for $O=M^x$ the symmetry generators commute with both the $O$  and the $e^{iH_1t}$ and we then obtain an expression analogous to Eq.~\eqref{eq:rho_0-trafo} with $\rho$ replaced by $M^x$. When $O=M^z_\pi$, $T_1$ and $R$ anticommute with the observable but commute with $H_1$. We therefore find, for a general element of the symmetry group $R^aT_1^b$ (where $a=0,1$ and $b=0,\dots,L-1$)~\footnote{We note in passing that the group generated by $R$ and $T_1$ is isomorphic to $D_L$, the dihedral group with $2L$ elements. This follows from $R^2=T_1^L=\mathbbm 1$ and the fact that $RT_1R=T_1^{-1}$. The latter equation can be checked by computing the action of both sides on a general computational basis element. These properties can be used to show that a general group element can be written in the form $R^aT_1^b$.}, that
\begin{align}
\label{eq:Mzpi-trafo}
    \braket{n|M^z_\pi(t)|m} &= (-1)^{a+b}\braket{n|(R^aT_1^b)^\dagger M^z_\pi(t)R^aT_1^b|m}\nonumber\\
    &=(-1)^{a+b}\braket{n'|M^z_\pi(t)|m'},
\end{align}
where now $\ket{m'}=R^aT_1^b\ket{m}$ and $\ket{n'}=R^aT_1^b\ket{n}$. Thus, from a single simulation yielding the matrix element $O_{nm}$, we can obtain at most $2L-1$ additional $O_{nm}$ between symmetry-related basis states at no cost.

\begin{figure*}[htb!]
\includegraphics[width=\linewidth]{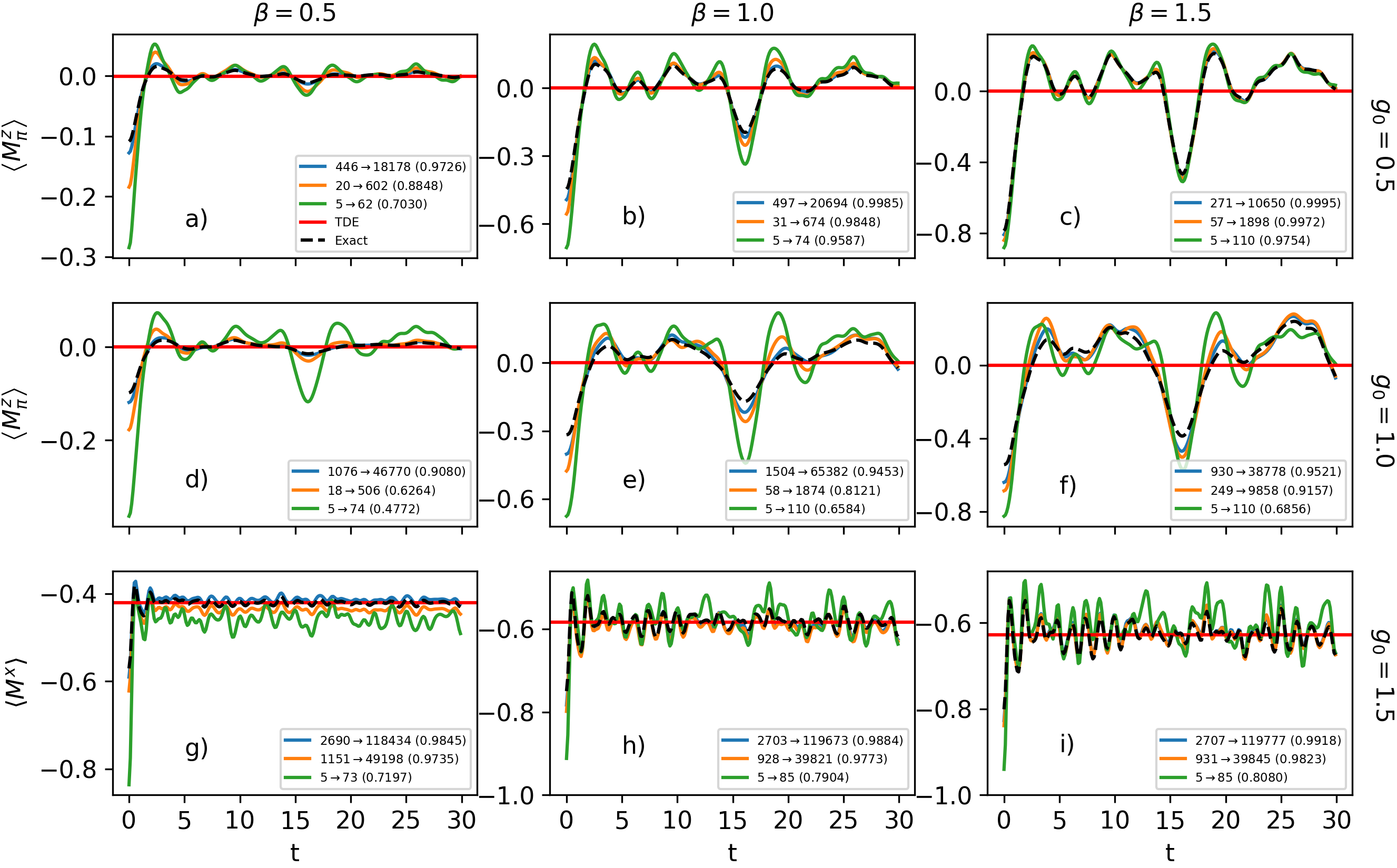}
\caption{
Thermal quench dynamics results for $L=12$ for various $\beta$ and $g_0$ (see labels at top and right, respectively). 
In each panel, the exact dynamics results calculated using ED are shown in black, and the red horizontal line indicates the TDE average of the observable, Eq.~\eqref{eq:TDE}.
Curves of different colors represent calculations using different $w$, resulting in truncated density matrices $\rho^w$ of different sizes $N_w$. To simulate the dynamics at a given $N_w$, $N_{\rm sim}$ simulations are performed; $N_{\rm sim}$ can be much less than $N_{w}$ due to the application of symmetries.
Each curve is labeled in the legend with the notation $N_{\rm sim}\rightarrow N_w\,(w)$.
}
\label{fig:dynamics_plots}
\end{figure*}

Finally, we note that Eq.~\eqref{eq:Mzpi-trafo} can be used to eliminate certain simulations entirely. In particular, if we are interested in $M^z_\pi(t)$ between eigenstates of $R$, we can apply Eq.~\eqref{eq:Mzpi-trafo} with $a=1,b=0$ to observe that, since $\ket{m'}=\ket{m}$ and $\ket{n'}=\ket{n}$ when $\ket{m}$ and $\ket{n}$ are eigenstates, we must have that $\braket{n|M^z_\pi(t)|m}=-\braket{n|M^z_\pi(t)|m}=0$ for those states. This allows us to explicitly exclude $2^{L/2}(2^{L/2}+1)/2$ $O_{nm}$ from Eq.~\eqref{eq:O(t)}, corresponding to the upper triangle of the matrix $O_{nm}$ in the space of reflection-symmetric states. Furthermore, since $T_1$ commutes with $H_1$ and anticommutes with $M^z_\pi$, we have that
\begin{align}
\label{eq:Mzpi_symex}
\braket{n|(T^b_1)^\dagger M^z_\pi(t)T_1^b|m} = (-1)^b\braket{n|M^z_\pi(t)|m}.
\end{align}
When $\ket m$ and $\ket n$ are $R$-eigenstates, these $O_{nm}=0$. This increases the number of excluded states by a multiplicative factor of at most $L$.

In summary, symmetries relate both $\rho_{mn}$ and $O_{nm}$. When $O$ transforms simply under the symmetry group of $H_1$ (e.g.~if it is invariant or acquires a minus sign), a single $O_{nm}$ yields a family of matrix elements $O_{n'm'}$ related by symmetry at no additional simulation cost. Furthermore, when $O_{nm}$ anticommutes with a symmetry generator (e.g. $M^z_\pi$), minus signs appear which can be used to additionally exclude matrix elements between any two eigenstates of that generator. Furthermore, $\rho$ and $O(t)$ are always Hermitian, so one need only consider upper-triangular matrix elements of both operators. Combining all of these simplifications allows us to reduce $N_{\rm sim}$, as we will demonstrate below.

\section{Simulation Results}
\label{Results}
We now present the results of classical simulations of thermal quench dynamics for the observables $M^z_\pi$ and $M^x$. We focus on two example systems, in Sec.~\ref{subsec:L12} a chain of length $L=12$ sites, and in Sec.~\ref{subsec:L16} a chain of length $L=16$. In the former case, exact numerical results are accessible via ED, so that detailed benchmarking can be performed as a function of $N_w$. In the latter case, full ED is impractical but DMQMC simulations can yield an accurate estimate of $\rho$ to simulate the $O(t)$ in Eq.~\eqref{eq:O(t)}. In this case, attention must be paid to the impact of the systematic error in the DMQMC algorithm on the subsequent dynamics.

\subsection{$L=12$-site Chain (ED Initial State)}
\label{subsec:L12}

For $L=12$ chains, $\rho = e^{-\beta H_0}/\text{Tr}(e^{-\beta H_0})$ can be obtained from $H_0$ via ED. An exact 
%noise-free 
simulation of the dynamics of an operator $O$ can then be obtained directly by computing the Heisenberg operator $O(t)=e^{iH_1t}Oe^{-iH_1t}$ where  $e^{-iH_1t}$ is obtained by direct matrix exponentiation using eigenstates of $H_1$. The late-time steady state value of $\braket{O(t)}_{\rho}$ can be calulated via the \emph{thermal diagonal ensemble} (TDE) average of $O$. This TDE average is defined by analogy with the diagonal ensemble (DE) average~\cite{nature06838} in pure state dynamics as
\begin{align}
\label{eq:TDE}
    \lim_{t\to \infty} \text{Tr}\bigl[ \rho O(t) \bigr]  =\frac{1}{Z_0} \sum_{E_0, E_1}e^{-\beta E_0} |\braket{E_0|E_1}|^2\braket{E_1|O|E_1},
\end{align}
where $E_0$ and $E_1$ are eigenvalues labeling eigenstates of $H_0$ and $H_1$, respectively. The TDE average can be viewed as a Boltzmann-weighted average of the DE value of $O$ for pure-state quenches from every eigenstate of $H_0$. We will say that the system has equilibrated when the time-average of $\braket{O(t)}_{\rho_0}$ reaches the TDE value.

In Fig.~\ref{fig:dynamics_plots}, we compare the exact result for $\braket{O(t)}_{\rho}$ to results obtained by evaluating Eq.~\eqref{eq:O(t)} using  $\rho^w$.
To calculate the contribution of the $N_w$ matrix elements to the dynamics, we perform $N_{\rm sim}$ simulations to obtain the corresponding matrix elements $O_{nm}$. [We count $O_{nm}$ as one simulation, although multiple simulations may be required using the superposition states~\eqref{eq:purepsiphi}.] Many $O_{nm}$ are related by symmetry, so $N_{\rm sim}\leq N_w$, often much less. In Fig.~\ref{fig:dynamics_plots}, we explore how varying $N_w$ affects accuracy.

\begin{figure}[t!]
\centering
\includegraphics[width=\linewidth]{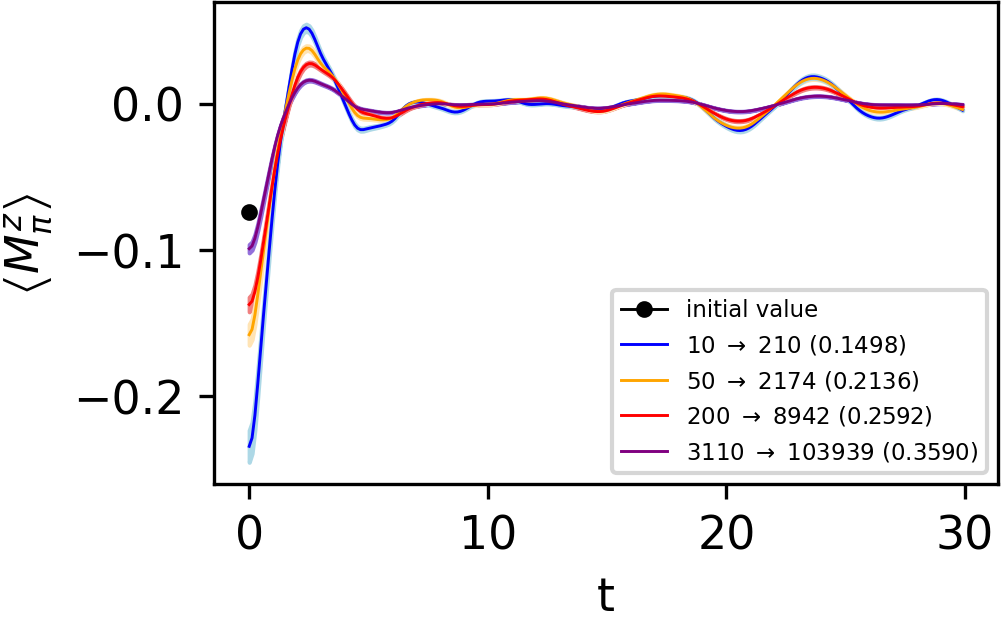}
 \caption{Quench dynamics for $L=16$, $\beta=0.5$, $g_0=1.0$, and $h_0=0.0$ using $\tilde{\rho}^w$ obtained from the DMQMC estimate $\tilde \rho$. The black dot denotes the initial value of $\left \langle M_{\pi}^z \right \rangle$ as calculated from the diagonal elements of $\tilde{\rho}$.
 The curves have confidence bands indicating the time-dependent value of the (statistical) error $\Delta \braket{O(t)}_{\tilde{\rho}}$, Eq.~\eqref{eq:dynamics_statistical_error}. Note that the error band does not overlap with the initial value due to the presence of truncation error. Curves are labeled as in Fig.~\ref{fig:dynamics_plots} and $w$ is computed with respect to $\tilde{\rho}$ .}
\label{fig:L16_DMQMC}
\end{figure}

%\begin{figure}[t!]
%\centering
%\includegraphics[width=\linewidth]{L16_DMQMC.png}
% \caption{Quench dynamics for $L=16$, $\beta=0.5$, $g_0=1.0$, and $h_0=0.0$ using $\tilde{\rho}^w$ obtained from the DMQMC estimate $\tilde \rho$. The black dot denotes the initial value of $\left \langle M_{\pi}^z \right \rangle$ as calculated from the diagonal elements of $\tilde{\rho}$.
% The blue and purple curves corresponding, respectively, to the smallest and largest $N_{\rm sim}$ have light blue and light purple confidence bands indicating the time-dependent value of the (statistical) error $\Delta \braket{O(t)}_{\tilde{\rho}}$, Eq.~\eqref{eq:dynamics_statistical_error}. Note that the error band does not overlap with the initial value due to the presence of truncation error. Curves are labeled as in Fig.~\ref{fig:dynamics_plots} and $w$ is computed with respect to $\tilde{\rho}$ .}
%\label{fig:L16_DMQMC}
%\end{figure}

The observable dynamics in Fig.~\ref{fig:dynamics_plots} is indicative of the initial and final locations of the quench. 
If the quench begins in the AFM phase [panels a)--c) and d)--f)] we use $O=M_{\pi}^z$ and perform dynamics simulations in the $z$-basis; if it begins in the PM phase [panels g)--i)] we use $O=M^x$ and simulate in the $x$-basis. 
Correlating the sampling basis with the observable in this way minimizes the $N_w$ needed to capture the expectation value at time $t=0$.
The TDE average of the observable, which indicates the late-time value of $\braket{O(t)}_{\rho}$, is observed to be nonzero only for quenches within the PM region ($g_0=1.5$). 
For $g_0=0.5$ [panels a)--c)], increasing $\beta$ generally results in larger $w$ captured for a fixed $N_{\rm sim}$.
This matches the expectation based on Fig.~\ref{fig:weight_plots}a), where the largest-$\beta$ $\rho^w$ required the smallest $N_w$. 
For $g_0=1.0$ [panels d)--f)], the initial state is at the $L\to\infty$ quantum critical point, but evidently retains some residual AFM order at finite size.
In this case, a larger $N_{\rm sim}$ is required to capture the dynamics as compared to the $g_0 = 0.5$ results.
Nevertheless, we still observe the general trend that increasing $\beta$ increases the $w$ captured by a fixed $N_{\rm sim}$.
Finally, for $g_0=1.5$ [panels g)--i)], we can only exclude simulations based on the smaller set of symmetries of $M^x$.
However, even after accounting for symmetries, we observe that $N_{\rm sim}$ needs to be about an order of magnitude larger to capture the dynamics than for $g_0=0.5$.
This is likely due to the fact that the spins polarize in the $x$-$z$ plane rather than purely along the $x$ axis when $g=h=1$.
Nevertheless, we observe in all cases that exploiting symmetries reduces $N_{\rm sim}$ by one to two orders of magnitude compared to $N_w$.

\subsection{$L=16$-site Chain (DMQMC Initial State)}
\label{subsec:L16}

We simulate thermal quench dynamics for $L=16$ using DMQMC to obtain $
\tilde{\rho}$ [see Eq.~\eqref{eq:rho_dmqmc}] for parameters $g_0=1, h_0=0, \beta=0.5$. The approximate initial value $\langle M^z_\pi(0)\rangle_{\tilde{\rho}^w}$ clearly approaches this DMQMC value as a function of $w$.
Additionally, we see oscillations about $0$ as expected for a quench from the quantum critical point into the PM phase, which suggests that the dynamics have equilibrated. These results further demonstrate the utility of symmetries, which allow $N_{\rm sim}\ll N_w$.

\begin{figure}[tb!]
\includegraphics[width=\linewidth]{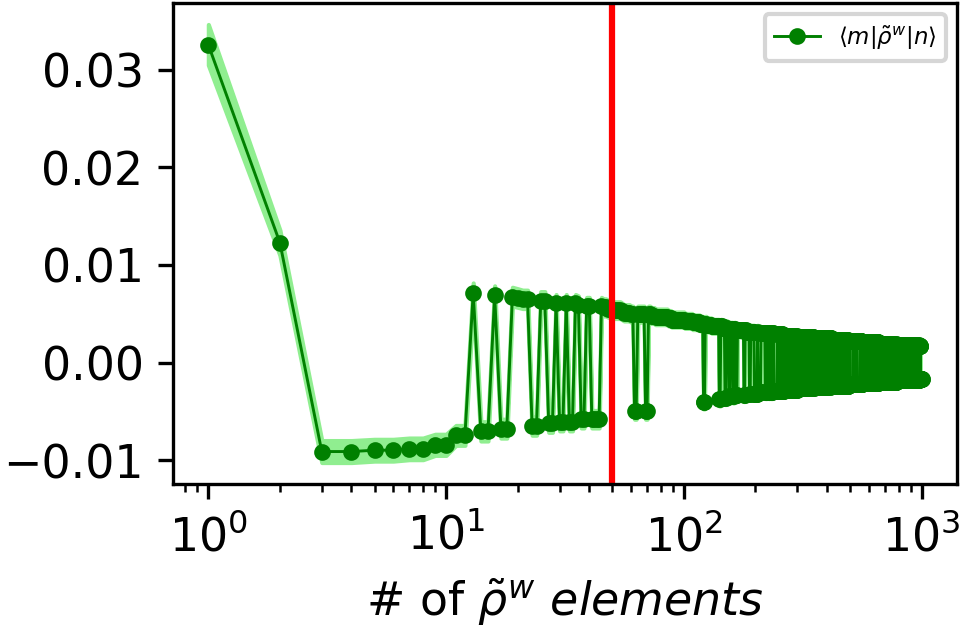}
\caption{Density matrix elements $\tilde{\rho}^w_{mn}$ for $L=16$, $\beta=0.5$, $g_0=1.0$, and $h_0=0.0$, plotted in order of decreasing magnitude. The sign of the density matrix elements becomes more incoherent as the elements become smaller. An $N_{\rm sim}$ of 50, denoted by the red line, produces 2174 observable elements through symmetries of $H_1$, corresponding to approximately $w=0.2136$ of $\tilde{\rho}^w$. Each point has an error $\Delta \tilde{\rho}^w_{mn}$  [Eqs.~\eqref{eq:rho_tilde_error_offdiag} and \eqref{eq:rho_tilde_error_diag}] indicated by the light green confidence band.}
\label{fig:dm_obs}
\end{figure}

The curves in Fig.~\ref{fig:L16_DMQMC} are shown with confidence bands indicating how the statistical uncertainty from finite $N_{\rm psip}$ in $\tilde{\rho}_{mn}$ propagates into the dynamics calculation. The detailed derivation of this error propagation is presented in Appendix~\ref{sec:statistical_error}.
We note that the uncertainty is time-dependent due to variations in the relative contributions of the $\braket{n|O(t)|m}$ as derived in Eq.~\eqref{eq:dynamics_statistical_error}. Of particular note is that the statistical error in the dynamics scales down with increasing $N_{\rm sim}$.

The dynamics plots in Fig.~\ref{fig:L16_DMQMC} are calculated using truncated density matrices $\tilde\rho^w$ capturing a weight of at most $w\approx 0.36$ of $\tilde\rho$. To understand why such a small weight might be sufficient,~Fig.~\ref{fig:dm_obs} shows $\tilde{\rho}_{mn}^w$ for the first $1000$ matrix elements sorted by magnitude from highest to lowest. We see that the matrix elements rapidly become very small, however care must be taken since their smallness can in principle be counteracted by their exponentially large number. Nevertheless, we also observe that, as the density matrix elements become smaller in magnitude, they begin to oscillate in sign and therefore contribute incoherently to Eq.~\eqref{eq:O(t)}. Indeed, such oscillations become more likely as the density matrix elements become smaller and the statistics noisier. (This is an example of a sign problem in Monte Carlo analysis.) This sign incoherence motivates our choice of a relatively small $w$ in our simulations and suggests that the simulation cost may scale favorably as a function of $L$.

\section{Discussion and Outlook}
\label{Discussion_and_outlook}

The E$\rho$OQ algorithm~\cite{Lamm:2018siq} provides an approach to simulating quench dynamics from a thermal initial state using a combination of classical and quantum techniques. 
The initial step of thermal state preparation is circumvented using a classical stochastic algorithm, while the classically hard task of time evolution is carried out on a quantum computer.
The desired operator dynamics is then reconstructed by a weighted average of the basis state dynamics results.
A crucial shortcoming of this approach is that, generically, the $N_{\rm sim}$ needed to reconstruct the thermal quench dynamics grows exponentially with system size.
In this work we have investigated the possibility of mitigating this issue by systematically truncating the initial density matrix.
We have also shown that symmetries can be exploited to further reduce the number of dynamics simulations needed to account for a fixed number of density matrix elements.

The efficacy of the methods developed here depends strongly on the initial temperature and Hamiltonian parameters.
For example, at low initial temperature and for initial Hamiltonian parameters for which the zero-temperature ground state is gapped, we find that the thermal quench dynamics of certain operators can be reconstructed from only a handful of pure state dynamics simulations [see Fig.~\ref{fig:dynamics_plots}c) and i)].
However, at high temperatures or for initial Hamiltonian parameters that correspond to a zero-temperature quantum critical point, substantially more dynamics simulations are required.
Regardless of the parameters of the initial state, we find that leveraging symmetries allows for a reduction by one to two orders of magnitude in the number of pure state dynamics simulations needed to capture the contribution from a fixed number of density matrix elements.
Although these techniques do not eliminate the exponential scaling issue, we expect that they will be indispensable for future implementations of E$\rho$OQ at system sizes comparable to or beyond those accessible to ED.

% Discuss issue of quantifying truncation error/bounding its effects on the dynamics as a direction for future work

\begin{figure}[t!]
\centering
\includegraphics[width=\linewidth]{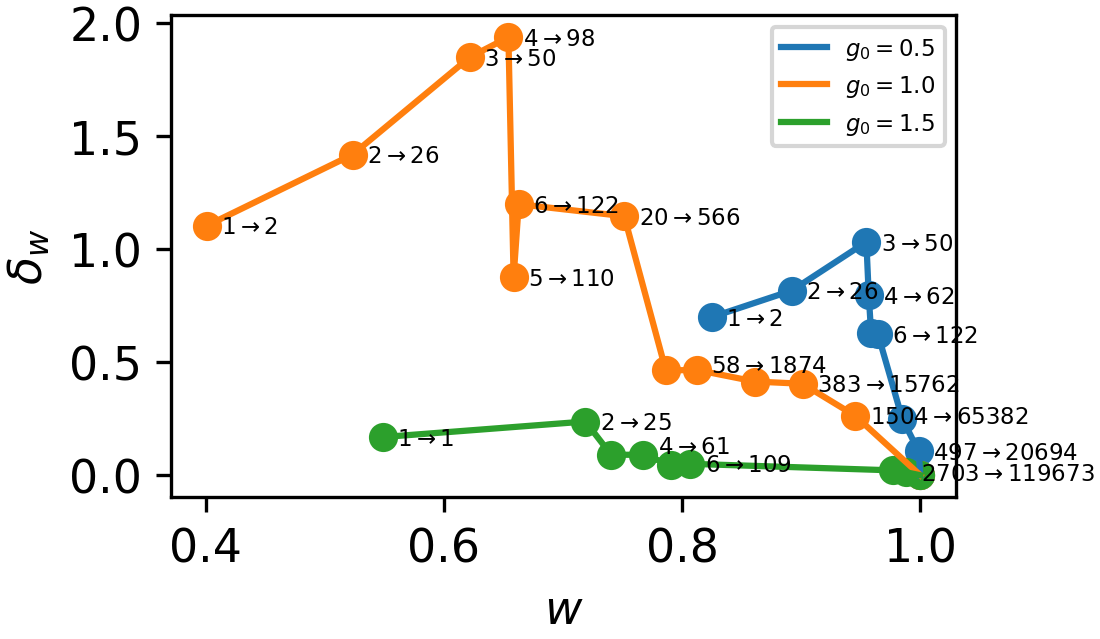}
 \caption{Dependence of $\delta_w$ [Eq.~\eqref{eq:trunc}] on $w$ for $L=12$, $\beta=1.0$, $h=1.0$, $h_0=0.0$, $g=1.0$, and $g_0=0.5,1.0,1.5$ (see legend). When $g_0=0.5$ or $1.0$, $O = M_{\pi}^z$ and when $g_0=1.5$, $O = M^x$. Some points are labeled with the notation $N_{\rm sim} \rightarrow N_w$.}
\label{fig:truncation_error}
\end{figure}

One challenge worth addressing in future work is the difficulty of quantifying the influence of density matrix truncation error on the dynamics of observables. In this work, we probed the effect of truncation on the dynamics by changing $N_{\rm sim}$, but it would be desirable to have an estimate of the statistical uncertainty in a dynamical expectation value due to truncation. This would allow one to estimate, from results with a fixed $N_{\rm sim}$, a confidence band around the simulated time trace of the observable. Such an understanding would become increasingly valuable as the simulations are scaled up to larger system sizes, to the point where the ``exact" dynamics from the DMQMC initial state is no longer simulable, like in the case of $L=16$ discussed above. With access to exact dynamics, it is possible to calculate truncation error. Fig.~\ref{fig:truncation_error} shows the complicated convergence of the truncation error defined as 
\begin{equation}
\label{eq:trunc}
    \delta_w=\frac{\sqrt{\frac{1}{T} \int_0^T dt | \left \langle O(t) \right \rangle_{\rho} - \left \langle O(t) \right \rangle_{\rho^w} |^2}}{\sqrt{\frac{1}{T} \int_0^T dt | \left \langle O(t) \right \rangle_{\rho} |^2}},
\end{equation}
with increasing weight $w$ and for $L=12$. The $\delta_w$ at fixed $N_{\rm sim}$ (see, e.g., the points corresponding to $N_{\rm sim}=4$) is largest for $g_0=1.0$,where more states contribute significantly to the dynamics.

% Discuss future applications: Thermal Green's functions, effects of initial temperature on dynamical phase transitions

A number of future applications of the techniques developed here can be envisioned. One example is the calculation of thermal Green's functions, which hinge on calculating quantities of the form $\braket{A(t)B(0)}_{\rho}=\text{Tr}(\rho A(t)B(0))$. Here the initial state and time evolution are defined with respect to the same Hamiltonian $H$, unlike the thermal quench protocols considered here. Nevertheless, the same density matrix truncation and symmetry approaches considered here could readily be applied. We also note that pure-state two-time correlation functions can be simulated on quantum computers with only constant overhead~\cite{PhysRevResearch.1.013006}. Another direction is to apply these methods to the simulation of dynamical phase transitions~\cite{Heyl2013,Heyl2018}, which have also been studied in the context of mixed-state dynamics~\cite{PhysRevB.96.180303,PhysRevB.96.180304}. It would be interesting to consider the fate of the zero-temperature dynamical phase transition~\cite{Heyl2013} when the initial state is at finite $T$. Simulating such phenomena requires calculating the Loschmidt echo, which can be achieved on quantum computers using Hadamard-test protocols~\cite{Ortiz01,Somma02} or ancilla-free versions thereof~\cite{PhysRevResearch.1.013006}. It may also be possible to incorporate some of the techniques considered here into classical tensor network algorithms for calculating finite temperature properties, such as the minimally entangled typical thermal states method~\cite{Stoudenmire_2010, PhysRevB.95.195148, PhysRevB.92.115105}.

\begin{acknowledgements}
This material is based on work supported by the U.S. Department of Energy, Office of Science, National Quantum Information Science Research Centers, Superconducting Quantum Materials and Systems Center (SQMS) under contract number DE-AC02-07CH11359. Fermilab is operated by Fermi Research Alliance, LLC under contract number DE-AC02-07CH11359 with the United States Department of Energy. T.I. acknowledges the Aspen Center for Physics, which is supported by National Science Foundation grant PHY-2210452, where part of this work was performed. P.P.O. thanks A. Vishwanath and Harvard University for hospitality during the final stages of this project. 
\end{acknowledgements}

\begin{appendix}

% \section{On generating free simulations with non-commuting $T_1$ and $R$}
% Suppose we have a quantum state defined by the bit string $\left| b_1 b_2 ... b_L \right>$ where $b_i \in \{0,1\}$. Let $R \in \mathbb{R}$ and $T_1 \in \mathbb{T}_1$ be mirror and one-site translation operators and the groups they belong in respectively. $\mathbb{R}$ and $\mathbb{T}_1$ are cyclic groups as they are each generated by a single element $R$ and $T_1$ such that $R^2=T_1^L=e$ where $e$ is the identity element. 

% \begin{align}
% R\left| b_1 b_2 ... b_L \right> = \left| b_L b_{L-1} ... b_1 \right>, \notag\\
% T_1R\left| b_1 b_2 ... b_L \right> = \left| b_1 b_L ... b_2 \right>,  \notag\\
% RT_1R\left| b_1 b_2 ... b_L \right> = \left| b_2 b_3 ... b_1 \right>,  \\
% T_1^{-1}\left| b_1 b_2 ... b_L \right> = \left| b_2 b_3 ... b_1 \right> 
% \end{align}
% from (A1) and (A2), $T_1^{-1} = RT_1R$ which using $R^2=e$ generalizes to $T_1^{-a} = RT_1^{a}R$ for some integer $a \in \{0, L-1\}$. Similarly 

% \begin{align}
% T_1\left| b_1 b_2 ... b_L \right> = \left| b_L b_1 ... b_{L-1} \right>, \notag\\
% RT_1\left| b_1 b_2 ... b_L \right> = \left| b_{L-1} b_{L-2} ... b_L \right>,  \notag\\
% T_1RT_1\left| b_1 b_2 ... b_L \right> = \left| b_L b_{L-1} ... b_1 \right>, \\
% R\left| b_1 b_2 ... b_L \right> = \left| b_L b_{L-1} ... b_1 \right> 
% \end{align}
% from (A3) and (A4), $T_1RT_1 = R$. $\mathbb{T}_1$ is a normal subgroup $\mathbb{T}_1 \triangleleft G$ and $G$ is the non-abelian semi-direct product group $G = \mathbb{T}_1 \rtimes \mathbb{R}$ with two generators such as $R$ and $T_1$ and cardinality $2L$.

\section{Statistical error}
\label{sec:statistical_error}
For a function $f(x_1, ..., x_n)$ with variables $x_1, ..., x_n$, the error is given by $\Delta f = \sqrt{\sum_{i}\left(  \frac{\partial f}{\partial x_i} \Delta x_i \right)^2}$. Consider $\braket{O(t)}_{\tilde{\rho}^w}= \sum_{m,n} \tilde{\rho}^w_{mn}  O_{nm}$.
Here 
\begin{equation}
    \tilde{\rho}^w_{mn} = \begin{cases}
    \frac{\chi_{mn}}{\chi_{\rm diag}^w} & |\tilde\rho_{mn}|>\epsilon\\
    0 & \mathrm{otherwise}
    \end{cases}
\end{equation}
where $\chi_{\rm diag}^w =  \sum_{(i,i) \in \mathcal W} \chi_{ii}$ where $\mathcal W$ is the set of non-zero matrix elements of $\tilde{\rho}^w$ (see Sec.~\ref{structure_and_truncation_of_the_initial_density_matrix}; recall that in practice we take $\chi_{mn}$ to be real.). All three, $\rho,\tilde{\rho},$ and $\tilde{\rho}^w$ are square and symmetric. We write $\chi_{mn}=\sigma_{mn}N_{mn}$ where $N_{mn}$ is the number of psips corresponding to the matrix element $\tilde\rho_{mn}$ and $\sigma_{mn}=\pm 1$ is the sign of the corresponding element, determined by the net charge of the psips. We also define the number of diagonal psips, $N^w_{\text{diag}} = \sum_{(i,i)\in\mathcal W}|\chi_{ii}| = \sum_{(i,i)\in\mathcal W}N_{ii}$. Poisson statistics implies that $\Delta N_{mn}^2=N_{mn}$. The error in the dynamics of the observable is given by
\begin{align}
\label{eq:dynamics_statistical_error}
\Delta \braket{O(t)}_{\tilde{\rho}^w} = \sqrt{ \sum_{m,n} (\Delta \tilde{\rho}^w_{mn}O_{nm})^2  }. 
\end{align}
When $m \neq n$, the error in $\tilde{\rho}_{mn}^w$ is 
\begin{align}
\label{eq:rho_tilde_error_offdiag}
\Delta \tilde{\rho}_{mn}^w &= \sqrt{ \left ( 
\frac{\partial \tilde{\rho}_{mn}^w}{\partial N_{mn}} \Delta N_{mn} \right )^2 + \sum_{i=1}\left (  \frac{ \partial \tilde{\rho}_{mn}^w}{\partial N_{ii} } \Delta N_{ii}  \right )^2 } \notag\\
%&= \sqrt{  \frac{N_{mn}}{(\chi^{w}_{\rm diag})^2} + \frac{N_{mn}^2}{(\chi_{\rm diag}^{w})^4}\sum_{i}N_{ii}  }\notag\\
&= \frac{\sqrt{N_{mn}}}{|\chi^w_{\rm diag}|}\sqrt{  1 + \frac{N_{mn}}{(\chi^{w}_{\rm diag})^2}N^w_{\rm diag} }.
\end{align}
When $m=n$, the error in $\tilde{\rho}_{mm}^w$ is 
\begin{align}
\label{eq:rho_tilde_error_diag}
\Delta \tilde{\rho}_{mm}^w &= \sqrt{  \sum_{i=1}\left (  \frac{ \partial \tilde{\rho}_{ mm}^w}{\partial N_{ii} } \Delta N_{ii}  \right )^2 } \notag\\
%&= \sqrt{\left( \frac{\sigma_{mm}\chi_{\rm diag}^w - N_{mm}}{(\chi^{w}_{\rm diag})^2} \right)^2N_{mm} + \left( \frac{N_{mm}}{(\chi_{\rm diag}^{w})^2}\right)^2 \sum_{i \neq m}N_{ii}} \notag\\
&=\frac{\sqrt{N_{mm}}}{|\chi^w_{\rm diag}|}\sqrt{1 - \frac{2 \chi_{mm}}{\chi^w_{\rm diag}} + \frac{N_{mm}}{(\chi^{w}_{\rm diag})^2}N^w_{\rm diag}}.
\end{align}

\section{Calculation of observable matrix elements}
\label{sec:computation_thermal_loschmidt_echo}
Here we discuss two ways to calculate the time-dependent matrix elements $O_{nm}$ on a quantum computer. For $n=m$, the calculation is straightforwardly accomplished by a direct measurement of $O$ following the dynamics under $H_1$ from the initial pure state $\ket{n}$. For $n \neq m$, and noticing that it is easier to compute diagonal expectation values $\braket{\psi|O(t) | \psi}$ than off-diagonal overlaps $\braket{\psi'|O(t) | \psi}$ where $\ket{\psi'} \neq \ket{\psi}$, we can perform time-evolution starting from $\ket{\psi^{\pm}_{nm}}$ and $\ket{\phi^{\pm}_{nm}}$ defined in Eq.~(\ref{eq:purepsiphi}). Efficient quantum circuits to prepare these states are discussed in the next Section~\ref{sec:initial_state_preparation}.
Computing the four diagonal expectation values yields
\begin{align}
    \braket{\psi^{\pm}_{nm} | O(t)| \psi^{\pm}_{nm}} &=\notag\\ \frac12 \Bigl[ O_{nn} +& O_{mm} \pm ( O_{nm} + O_{mn} ) \Bigr]\\
    \braket{\phi^{\pm}_{nm} | O(t) | \phi^{\pm}_{nm}} &=\notag\\ \frac12 \Bigl[ O_{nn} +& O_{mm} \pm i (O_{nm}  - O_{mn} ) \Bigr]\,.
\end{align}
If $O$ is Hermitian, i.e. $O^\dag = O \implies O_{mn} = O_{nm}^*$, both expectation values are real:
\begin{align}
    \braket{\psi^{\pm}_{nm} | O(t) | \psi^{\pm}_{nm}}  &=  \frac12 \Bigl( O_{nn} + O_{mm} \pm 2 \text{Re} \, O_{nm} \Bigr)\\
    \braket{\phi^{\pm}_{nm} | O(t) | \phi^{\pm}_{nm}} &= \frac12 \Bigl( O_{nn} + O_{mm} \mp 2 \text{Im} \, O_{nm} \Bigr) \,.
\end{align}
Eq.~\eqref{eq:Onm} follows directly from the above.

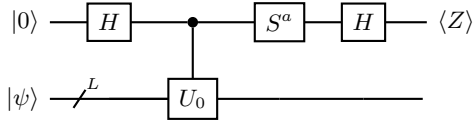
\begin{figure}[t    !]
\begin{quantikz}
\lstick{$\ket{0}$} & \gate{H} & 
\ctrl{1} &  \gate{S^a} & \gate{H} &  \rstick{$\left \langle Z \right \rangle$} \qw \\
\lstick{$\ket{\psi}$} &  \qwbundle{L} &  \gate{U_0} \qw&  \qw & \qw & \qw \ 
\end{quantikz}
 \caption{Hadamard test circuit determining $\text{Re}\braket{\psi|U_0|\psi}$ and $\text{Im}\braket{\psi|U_0|\psi}$ when $a=0$ and $a=1$ respectively.}
\label{fig:hadamard_test}
\end{figure}

Another method to measure off-diagonal matrix elements of observables is the Hadamard test~\cite{Ortiz01,Somma02}, whose circuit is shown in Fig.~\ref{fig:hadamard_test}.
The Hadamard test can be applied to measure off-diagonal elements of a unitary operator $U_0$.
We can use it to measure $\braket{n|O(t)|m}$ by expanding the Hermitian operator $O$ in the Pauli basis, $O=\sum_\alpha P_\alpha$ where $P_\alpha$ are Pauli strings.
If $\ket{m}$ and $\ket{n}$ are related by $\ket{m}=\prod_{j\in \mathcal S_{mn}}X_j\ket{n}$, where $\mathcal S_{mn}$ is the set of sites where the computational basis states $\ket{m}$ and $\ket{n}$ differ, then we can write
\begin{align}
    \braket{n|P_\alpha(t)|m} = \braket{n|e^{iH_1t} P_\alpha e^{-iH_1t} \prod_{j\in \mathcal S_{mn}}X_j|n}.
\end{align}
We can define $\ket{\psi}=\ket{n}$ and $U_0=e^{iH_1t} P_\alpha e^{-iH_1t} \prod_{j\in \mathcal S_{mn}}X_j$ in Fig.~\ref{fig:hadamard_test} and then sum over $\alpha$ to obtain the desired result.
The Hadamard test is unwieldy because it requires a controlled-$U_0$ gate controlled by an ancilla qubit.
As an alternative to the Hadamard test, one can also explore protocols like the ones proposed in Ref.~\cite{PhysRevResearch.1.013006} that avoid the need for an ancilla-controlled $U_0$ gate at the expense of running more direct measurement circuits.
Whether such methods are more desirable than that of preparing the superposition states \eqref{eq:purepsiphi} is hardware-dependent.
% Alternatively we can find off-diagonal elements by initializing $\ket{\psi} = \ket{m}$ and incorporating $X$-gates onto a unitary gate $U_0$ to find the real and imaginary values of the appropriate off-diagonal element $\braket{m| \Pi_j X_j U_0|m} = \braket{n|U_0|m}$ where the $j$ label all the qubits that are flipped and $U = \Pi_j X_j U_0$ is still unitary. Since the space of observables in the complex 2 dimensional Hilbert Space is spanned by Pauli matrices, any time-evolved observable in this space can be broken down into a sum of Pauli matrices which are unitary thus can be used in the Hadamard test. One downside of this approach is the decomposition of a controlled unitary as shown in Fig.~\ref{fig:hadamard_test} in terms of native gates can require many of them. In finding real and imaginary parts of off-diagonal elements following Ref.~\cite{PhysRevResearch.1.013006}, 8 circuits are needed -- 4 for each superposition state, but without an ancilla. For computing thermal correlation functions $\braket{A(t)B(0)}_{\rho}=\sum_{mn}\rho_{mn}\braket{n|A(t)B|m}$, each element $\braket{n|A(t)B|m}$ from the trace is computed using Eq.~(3) in \cite{PhysRevResearch.1.013006}, by setting $W = e^{-iHt}$, $U=A$, and decomposing $B$ into a sum of Pauli strings represented by $G$ such that $G^2=I$, summing over all the terms involved then weighting each $\braket{n|A(t)B|m}$ with the corresponding density matrix weight $\rho_{mn}$ and finally summing to get $\braket{A(t)B(0)}_{\rho}$. 

\section{Initial superposition state preparation}
\label{sec:initial_state_preparation}

\begin{figure}[t]
\begin{quantikz}[column sep=4pt, row sep={20pt,between origins}]
\lstick{$\ket{0}$} & \gate{X^{a_1}}\slice{1} & \gate{H} \slice{2} & \gate{S^{a_0}} \slice{3}  & \ctrl{1} \slice{4} & \ctrl{2} &\qw \slice{5} & \ctrl{4} & \qw & \qw & \qw & \qw   \\
\lstick{$\ket{0}$} & \qw & \gate{X^{a_2}} & \qw  & \targ{} & \qw & \ctrl{2} & \qw & \ctrl{4} & \qw & \qw & \qw  \\
\lstick{$\ket{0}$} & \qw & \gate{X^{a_3}} & \qw & \qw & \targ{} & \qw & \qw & \qw & \ctrl{4} & \qw & \qw  \\
\lstick{$\ket{0}$} & \qw & \gate{X^{a_4}} & \qw & \qw & \qw & \targ{} & \qw & \qw & \qw & \ctrl{4} & \qw  \\
\lstick{$\ket{0}$} & \qw & \gate{X^{a_5}} & \qw & \qw & \qw & \qw & \targ{} & \qw & \qw & \qw & \qw \\
\lstick{$\ket{0}$} & \qw & \gate{X^{a_6}} & \qw & \qw & \qw & \qw & \qw & \targ{} & \qw & \qw & \qw \\
\lstick{$\ket{0}$} & \qw & \gate{X^{a_7}} & \qw & \qw & \qw & \qw & \qw & \qw & \targ{} & \qw & \qw \\
\lstick{$\ket{0}$} & \qw & \gate{X^{a_8}} & \qw & \qw & \qw & \qw & \qw & \qw & \qw & \targ{} & \qw
\end{quantikz}
 \caption{Circuit for preparing $\ket{\psi^\pm_{nm}}$ and $\ket{\phi^\pm_{nm}}$, Eq.~\eqref{eq:purepsiphi} for $\ket{m}=\prod^L_{j=1}X_j\ket{n}$, following Ref.~\cite{Cruz}. The $X$-gate powers $a_i \in \{0,1\}$ determine whether the $X$-gates are applied or not in preparing a specific state with $i \in \{1, ..., L\}$. The state     $\ket{\psi^{\pm}_{nm}}$ is given when $a_0 = 0$ and the state $\ket{\phi^{\pm}_{nm}}$ is given when $a_0=1$ by application of an S-gate. The diagram shows the state preparation for $L=8$ but generalizes to $L=2^k$ where $k$ is a positive integer. The red lines distinguish the circuit layers.}
 \label{fig:max_sup}
\end{figure}
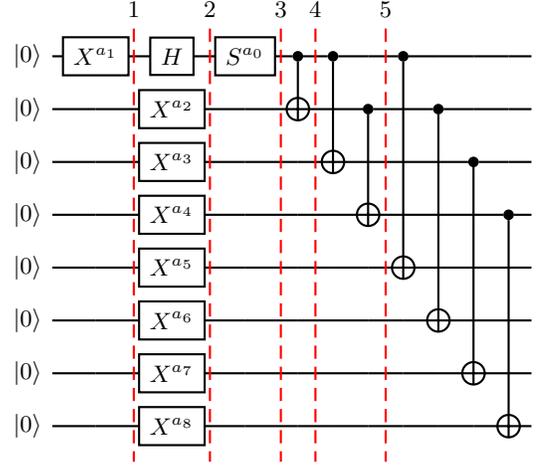

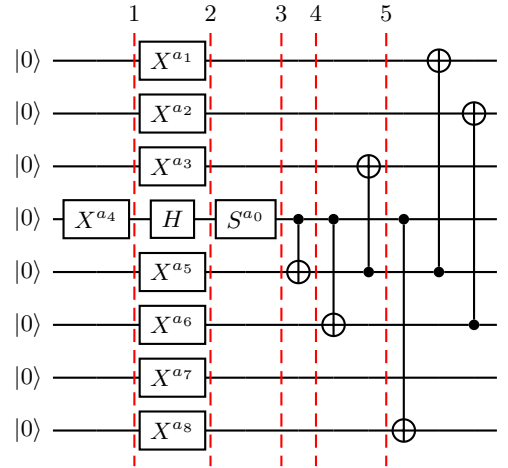
\begin{figure}[b]
\begin{quantikz}[column sep=4pt, row sep={20pt,between origins}]
\lstick{$\ket{0}$} & \qw\slice{1} & \gate{X^{a_1}} \slice{2} & \qw \slice{3} & \qw \slice{4} & \qw &\qw \slice{5} & \qw & \targ{} & \qw & \qw   \\
\lstick{$\ket{0}$} & \qw & \gate{X^{a_2}} & \qw & \qw & \qw & \qw & \qw & \qw & \targ{} & \qw  \\
\lstick{$\ket{0}$} & \qw & \gate{X^{a_3}} & \qw & \qw & \qw & \targ{} & \qw & \qw & \qw & \qw  \\
\lstick{$\ket{0}$} & \gate{X^{a_4}} & \gate{H} & \gate{S^{a_0}} & \ctrl{1} & \ctrl{2} & \qw & \ctrl{4} & \qw & \qw & \qw  \\
\lstick{$\ket{0}$} & \qw & \gate{X^{a_5}} & \qw & \targ{} & \qw & \ctrl{-2} & \qw & \ctrl{-4} & \qw & \qw \\
\lstick{$\ket{0}$} & \qw & \gate{X^{a_6}} & \qw & \qw & \targ{} & \qw & \qw & \qw & \ctrl{-4} & \qw \\
\lstick{$\ket{0}$} & \qw & \gate{X^{a_7}} & \qw & \qw & \qw & \qw & \qw & \qw & \qw & \qw \\
\lstick{$\ket{0}$} & \qw & \gate{X^{a_8}} & \qw & \qw & \qw & \qw & \targ{} & \qw & \qw & \qw
\end{quantikz}
 \caption{An example circuit, adapting the construction of Ref.~\cite{Cruz}, for preparing superposition states $\ket{\psi^\pm_{nm}}$ and $\ket{\phi^\pm_{nm}}$, Eq.~\eqref{eq:purepsiphi}, in cases where $\ket{n}$ and $\ket{m}$ differ on fewer than $L$ sites. When all $a_i = 0$, this gives the state $\frac{1}{\sqrt{2}}\left( \left | 0\right \rangle^{\otimes 8} + \left | 1\right \rangle^{\otimes 6}\otimes \ket{01} \right)$.}
 \label{fig:arb_sup}
\end{figure}

Here we consider the quantum resource cost of preparing the superposition states \eqref{eq:purepsiphi} on a quantum computer.
Preparing such states requires at most $O(L)$ CNOT gates: indeed, for $\ket{n}=\ket{0\dots0}$ and $\ket{m}=\ket{1\dots 1}$, the states \eqref{eq:purepsiphi} are variants of the GHZ state~\cite{Greenberger89}.
More generally, the complexity of state-preparation is highest for superpositions of states satisfying $\ket{m}=\prod^L_{j=1}X_j\ket{n}$, i.e. ones that differ on all sites.
The circuit depth needed to prepare such states can be reduced in certain cases by parallelizing the CNOT gates: Ref.~\cite{Cruz} found that the circuit depth can be reduced to $O(\ln L)$ assuming sufficient qubit connectivity and a system size that is a power of two. 
A general example of such a circuit for $L=8$ is shown in Fig.~\ref{fig:max_sup}. 
The $2^{L+1}$ superposition states $\ket{\psi^\pm_{nm}}$ and $\ket{\phi^\pm_{nm}}$ for which $\ket{m}=\prod^L_{j=1}X_j\ket{n}$ are indexed by $X$-gate powers $a_i \in \{0,1\}$, $i=0,\dots,L$. $a_0$ determines whether the state is of type $\ket{\psi}$ or $\ket{\phi}$, $a_1$ determines the $\pm$ sign, and the remainder dictate the pattern of bits in each state.
For example, setting all $a_i=0$ in Fig.~\ref{fig:max_sup} gives the GHZ state $\frac{1}{\sqrt{2}}\left( \left | 0\right \rangle^{\otimes 8} + \left | 1\right \rangle^{\otimes 8} \right)$ while setting  $a_2=1$ gives the GHZ-like states $\frac{1}{\sqrt{2}}\left( \left | 01010101 \right \rangle + \left | 10101010 \right \rangle \right)$.
Fewer CNOT gates are required for $\ket{m}$ and $\ket{n}$ that differ on fewer sites.
For a general off-diagonal element, $\ket{m}$ and $\ket{n}$ differ on at least one qubit.
Applying a Hadamard gate to one of these qubits, followed by a set of appropriately parallelized CNOTs, generates the desired superposition.
An example circuit is shown in Fig.~\ref{fig:arb_sup}.

\section{Thermal diagonal ensemble}
\label{sec:thermal_diagonal_ensemble}

We use the following expression for the TDE [$U(t) = e^{-iH_1t}$],
\begin{align}
\left \langle O(t) \right \rangle_\rho &= \frac{1}{Z_0}\text{Tr}\bigl[  e^{-\beta H_0} U^{\dagger}(t) O U(t) \bigr]\notag\\
&=\frac{1}{Z_0} \sum_{E_0} \braket{E_0| e^{-\beta H_0} e^{iH_1t} O e^{-iH_1t} |E_0}\\
=\frac{1}{Z_0} &\sum_{E_0,E_1,E'_1}e^{-\beta E_0}\braket{E_0|E_1}e^{i(E_1-E_1')t}\braket{E_1| O |E_1^{\prime}}\braket{E_1^{\prime}|E_0}\notag
\end{align}
where the sum is performed over $\ket{E_0}$ and $\ket{E_1},\ket{E_1'}$ are eigenstates of $H_0$ and $H_1$ respectively. Taking the infinite-time limit in analogy with Eq.~(2) from \cite{nature06838}, i.e., assuming attenuation of the off-diagonal terms due to temporal dephasing, we find
\begin{align}
\lim_{t\to \infty} \left \langle O(t) \right \rangle_\rho  =\frac{1}{Z_0} \sum_{E_0, E_1}e^{-\beta E_0} |\braket{E_0|E_1}|^2\braket{E_1|O|E_1}.
\end{align}
In the infinite-temperature limit, 
\begin{align}
\lim_{t\to \infty} \left \langle O(t) \right \rangle_\rho  =&\frac{1}{Z_0} \sum_{E_0, E_1} \braket{E_1|E_0}\braket{E_0|E_1}\braket{E_1|O|E_1}\notag\\
=&\frac{1}{Z_0} \sum_{E_1} \braket{E_1|E_1}\braket{E_1|O|E_1}\notag\\
=&\frac{1}{Z_0}  \text{Tr}\bigl[ O(t) \bigr] = \frac{1}{2^L}  \text{Tr}\bigl[ O(0) \bigr]
\end{align}
as expected. In the zero-temperature limit,
\begin{align}
\lim_{t\to \infty} \text{Tr}\bigl[ \rho O(t) \bigr] = \sum_{E_1}  |\braket{E_0^G|E_1}|^2\braket{E_1|O|E_1}
\end{align}
in analogy with Eq.~(2) from \cite{nature06838}. This corresponds to a pure-state quench from $\ket{E_0^G}$, the ground state of $H_0$.

\end{appendix}
\bigskip

\bibliography{refs}

\end{document}